\documentclass[twocolumn]{jpsj3} 
\usepackage{color}
\title{
Spin-Flop Phenomenon of Two-Dimensional Frustrated Antiferromagnets 
\\
without Anisotropy in Spin Space
}
\catcode`\@=11
\def\simle{\mathrel{\mathpalette\@versim<}}   
\def\simge{\mathrel{\mathpalette\@versim>}}   
\def\@versim#1#2{\lower2.5pt\vbox{\baselineskip0pt \lineskip-.5pt
   \ialign{$\m@th#1\hfil##\hfil$\crcr#2\crcr\sim\crcr}}}
\catcode`\@=12

\author{Hiroki Nakano$^{1}$
\thanks{E-mail: hnakano@sci.u-hyogo.ac.jp}, 
T\^oru Sakai$^{1,2}$
\thanks{E-mail: sakai@spring8.or.jp}, 
and 
Yasumasa Hasegawa$^{1}$ 
\thanks{E-mail: hasegawa@sci.u-hyogo.ac.jp}
}

\inst{$^{1}$Graduate School of Material Science, 
University of Hyogo,
Kamigori, 
Hyogo 678-1297, Japan \\
$^{2}$
Japan Atomic Energy Agency, SPring-8, 
Sayo, Hyogo 679-5148, Japan 
}

\recdate{\today}

\abst{
Motivated by a recent finding of a spin-flop phenomenon 
in a system without anisotropy in spin space reported 
in the $S=1/2$ Heisenberg antiferromagnet 
on the square-kagome lattice,  
we study the $S=1/2$ Heisenberg antiferromagnets 
on two other lattices composed of vertex-sharing triangles 
by the numerical diagonalization method.  
One is a novel lattice including a 
{\it shuriken} shape with four teeth; 
the other is 
the kagome lattice with $\sqrt{3}\times\sqrt{3}$-structure 
distortion, 
which includes a {\it shuriken} shape with six teeth. 
We find in the magnetization processes of these systems 
that a magnetization jump accompanied by a spin-flop phenomenon 
occurs at the higher-field-side edge of 
the magnetization plateau 
at one-third the height of saturation. 
This finding indicates that 
the spin-flop phenomenon found in the isotropic system 
on the square-kagome lattice is not an exceptional case. 
}

\begin{document}
\maketitle

\section{Introduction} 

In magnetization processes of magnetic materials, 
each material shows various behaviors as its characteristics. 
One of such macroscopic behaviors is the magnetization jump. 
There are several microscopic origins 
of this jump. 
As a trivial case, a ferromagnetic system shows 
a jump in its ground-state magnetization process 
when the external field is changed from negative to positive 
values in a fixed direction. 
The same type of jump in magnetization occurs 
in a ferrimagnetic system\cite{Sakai_Okamoto_ferri}. 
There is no change in lines along microscopic spin directions 
between the state under a negative field and 
that under a positive field. 
As another origin of the jump, 
a spin-flop phenomenon\cite{Neel_spin_flop,
kohno_aniso2D,sakai_aniso} is widely known; 
the phenomenon is the occurrence of 
an abrupt change in lines along microscopic spin directions 
while the states change owing to the increase in magnetic field. 
We distinguish the type of 
magnetization jump in the ferromagnet 
and ferrimagnet mentioned above 
from the spin-flop phenomenon. 
It is generally understood that this phenomenon occurs
when a certain anisotropy essentially exists in a magnetic system. 
One of the notable experimental realizations of such a spin-flop phenomenon 
is CsCuCl$_{3}$, whose magnetization jump is explained 
by the mechanism based on the spin-flop phenomenon 
due to the spin anisotropy of the system\cite{Nikuni_Shiba}.

\begin{figure}[htb]
\begin{center}
\includegraphics[width=7cm]{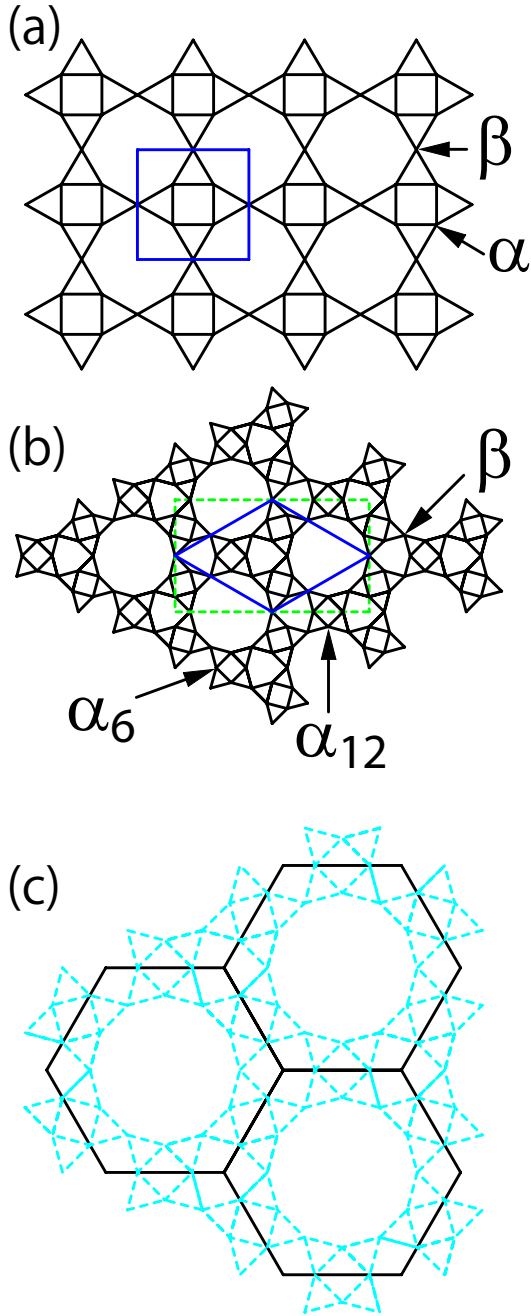}
\end{center}
\caption{(Color) (a) Square-kagome lattice and 
(b){\it shuriken}-bonded honeycomb lattice. 
Blue large square in (a) and 
blue rhombus in (b) represent 
a unit cell of each lattice. 
In (b), a rectangle surrounded by green dotted lines 
denotes a finite-size cluster of $N_{\rm s}=36$. 
(c) illustrates the relationship 
between the ordinary honeycomb lattice (black solid line) 
and the {\it shuriken}-bonded honeycomb lattice 
(light blue dotted line). 
}
\label{fig1}
\end{figure}

Under these circumstances, on the other hand, 
a recent investigation\cite{shuriken_lett} 
showed that a spin-flop phenomenon 
occurs in the $S=1/2$ Heisenberg antiferromagnet 
on the square-kagome (squagome) lattice 
shown in Fig.~\ref{fig1}(a) 
despite the fact that 
the system has no anisotropy in spin space. 
This finding became a counterexample 
to a general understanding of the spin-flop phenomenon. 
Unfortunately, 
it is unclear whether this behavior of the model 
is an exceptional and rare behavior of an irregular 
model or a general phenomenon that is observed 
in various systems sharing a spin-isotropic feature. 
Hereafter, the models that we consider in this study are limited 
to systems including Heisenberg-type interactions only. 

This lattice was originally 
introduced in Ref. \ref{squagome_Siddharthan_Georges} 
from the viewpoint of the relationship between 
the spin model on this lattice and the eight-vertex model. 
A numerical study of the antiferromagnet on this lattice 
was also reported in Refs. \ref{squagome_Tomczak} 
and \ref{squagome_Richter}. 
The lattice shares several characteristics 
with the kagome lattice; however, 
Ref.~\ref{shuriken_lett} showed that 
there is a distinct different behavior 
between the $T=0$ magnetization process 
of the antiferromagnet 
on the lattice shown in Fig.~\ref{fig1}(a) 
and that of the kagome-lattice antiferromagnet 
at approximately the one-third height of the saturation, 
particularly at the higher-field edge of this height. 
Therefore, the finding of the spin-flop phenomenon 
in the Heisenberg antiferromagnet on the square-kagome lattice 
requires us to reexamine our understanding of the mechanism 
underlying the spin-flop phenomenon. 

\begin{figure}[htb]
\begin{center}
\includegraphics[width=7cm]{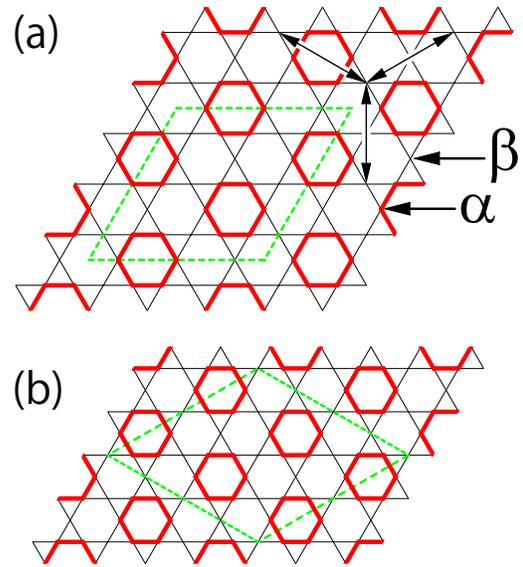}
\end{center}
\caption{(Color) 
The kagome lattice with distortion of 
$\sqrt{3}\times\sqrt{3}$ structure is illustrated 
by red thick and black thin solid lines. 
In (a) and (b), rhombuses surrounded by green dotted lines 
show finite-size clusters of $N_{\rm s}=27$ and 36, respectively. 
An arrow with a head at each of the two edges indicates 
the most distant pair in the finite-size cluster in (a). 
}
\label{fig2}
\end{figure}

Under the above mentioned circumstances, 
it is important to examine whether a spin-flop phenomenon 
also occurs in the spin-isotropic antiferromagnet 
on other lattices. 
If it is confirmed, one finds that  
a spin-flop phenomenon of spin-isotropic antiferromagnets 
is not a rare occurrence in 
the square-kagome-lattice antiferromagnet. 
In this paper, then, let us consider the lattices shown 
in Figs.~\ref{fig1}(b) and \ref{fig2}. 

The purpose of the present study 
is to present the second and third examples 
of the spin-flop phenomenon of spin-isotropic antiferromagnets.  
The examples are important because 
they possibly lead 
to clarifying the mechanism of the spin-flop phenomenon 
and to our deeper understanding of the frustration effect 
in quantum systems. 

The paper is organized as follows. 
In the next section, the model Hamiltonian and 
the calculation method are explained. 
The third section is devoted to the presentation 
and discussion of our main results. 
In the final section, 
the conclusion and future prospects are given. 

\section{Model Hamiltonians and Method} 
As mentioned above, we consider the lattices 
illustrated in Figs.~\ref{fig1}(b) and \ref{fig2}. 
In comparison with the situation that 
the square-kagome lattice shown in Fig.~\ref{fig1}(a) 
includes a {\it shuriken} structure 
formed from four equivalent regular triangles
in its unit cell, 
the lattice shown in Fig.~\ref{fig1}(b) also includes 
the same {\it shuriken} structure\cite{shuriken_lett}, 
but the network in the two-dimensional plane is different from 
that in the square-kagome lattice. 
As a consequence, the lattice shown in Fig.~\ref{fig1}(b) 
includes three {\it shurikens} in its unit cell. 
If one replaces each {\it shuriken} with a simple bond, 
one can obtain the honeycomb lattice [see Fig.~\ref{fig1}(c)]; 
therefore, let us call the lattice shown in Fig.~\ref{fig1}(b) 
the {\it shuriken}-bonded honeycomb (SBH) lattice hereafter,  
because the antiferromagnet on this lattice has not 
been investigated to the best of our knowledge. 
Note here that the SBH lattice is also related 
to the ordinary kagome lattice 
because the latter is reproduced from the SBH lattice 
when each {\it shuriken} in the SBH lattice is replaced 
by a simple vertex at the center of the {\it shuriken} 
and when one links all the nearest-neighbor pairs of new vertices 
after the replacement. 
In the SBH lattice, there are eighteen vertices in each unit cell. 
Vertices are divided into three groups:  
one is a vertex of the square inside a {\it shuriken} 
and a vertex of the hexagon,  
another is a vertex of the square 
and a vertex of the dodecagon, and 
the other is a vertex of the hexagon 
and a vertex of the dodecagon; 
hereafter, let us call the vertex sites in these groups 
the $\alpha_{6}$, $\alpha_{12}$, and $\beta$ sites, respectively. 
Note here that each vertex is shared by the neighboring 
two triangles; thus, we have the coordination number as 
$z=4$ in the SBH lattice; the characteristics 
are similar to those of the ordinary kagome lattice and the 
square-kagome lattice. 

On the other hand, the lattice in Figs.~\ref{fig2}(a) and 
\ref{fig2}(b) is obtained by a tuning 
of the interaction strength 
at the red thick bonds in the ordinary kagome lattice. 
The tuning leads to 
the so-called $\sqrt{3}\times\sqrt{3}$ distortion. 
Each unit cell includes nine vertices, 
six of them form a red thick hexagon. 
The hexagon is accompanied by six triangles 
in sharing their edges; 
the hexagon with the six triangles seem 
to form another type of {\it shuriken} 
with six teeth\cite{comment_shuri_six_teeth}, 
in contrast to the {\it shuriken} with four teeth 
in the SBH lattice shown in Fig.~\ref{fig1}(b). 
Owing to the distortion, vertices become divided 
into two groups: one is a vertex of a hexagon 
and one is not. 
Hereafter, let us call the vertex site in the former 
(latter) group the $\alpha$ ($\beta$) site. 
The magnetization process of the antiferromagnet 
on this $\sqrt{3}\times\sqrt{3}$-distorted kagome lattice
was first described in Ref.~\ref{Hida_kagome}, 
in which a similar backbending behavior 
in the finite-size magnetization process was observed 
at the higher-field edge of the one-third height 
of the saturation. 
Unfortunately, it was not able to examine 
whether the behavior is an artifact due the finite-size effect 
or a truly thermodynamic behavior 
within the limited system sizes treated in Ref. \ref{Hida_kagome}.  
In the present study, then, 
we carry out calculations of a larger cluster 
and obtain the magnetization process together with the information 
of local magnetizations, by which we tackle 
the backbending behavior of the antiferromagnet 
on the $\sqrt{3}\times\sqrt{3}$-distorted kagome lattice. 

The Hamiltonian that we study in this research is given by 
${\cal H}={\cal H}_0 + {\cal H}_{\rm Zeeman}$, where 
\begin{equation}
{\cal H}_0 = \sum_{\langle i,j\rangle} J 
\mbox{\boldmath $S$}_{i}\cdot\mbox{\boldmath $S$}_{j} , 
\label{H_shuriken_bond_honey}
\end{equation}
for the model on the lattice shown in Fig.~\ref{fig1}(b) 
and 
\begin{equation}
{\cal H}_0 = \sum_{\langle i,j\rangle \in {\rm black \ bonds}} 
J_{\rm A} 
\mbox{\boldmath $S$}_{i}\cdot\mbox{\boldmath $S$}_{j} 
+
\sum_{\langle i,j\rangle \in {\rm red \ bonds}} J_{\rm B} 
\mbox{\boldmath $S$}_{i}\cdot\mbox{\boldmath $S$}_{j} 
, 
\label{H_distorted_kagome}
\end{equation}
for the model on the distorted kagome lattice 
shown in Fig.~\ref{fig2}.  
Let us emphasize here that 
${\cal H}_{0}$ is isotropic in spin space. 
${\cal H}_{\rm Zeeman} $ is given by 
\begin{equation}
{\cal H}_{\rm Zeeman} = - h \sum_{j} S_{j}^{z} .  
\label{H_zeeman}
\end{equation}
Here, $\mbox{\boldmath $S$}_{i}$ 
denotes the $S=1/2$ spin operator 
at site $i$ illustrated 
by the vertex in Figs.~\ref{fig1} and \ref{fig2}. 
The sum of ${\cal H}_0$ runs over all the pairs 
of spin sites linked by solid lines 
in Figs.~\ref{fig1} and \ref{fig2}.  
Energies are measured in units of $J$ 
for the lattice shown in Fig.~\ref{fig1}(b) 
and $J_{\rm A}$ for the distorted kagome lattice; 
hereafter, we set $J=1$ and $J_{\rm A}=1$.  
The number of spin sites is denoted by $N_{\rm s}$. 
We impose the periodic boundary condition 
for clusters with site $N_{\rm s}$, 
which are shown in Figs.~\ref{fig1}(b), \ref{fig2}(a), 
and \ref{fig2}(b). 

We calculate the lowest energy of ${\cal H}_0$ 
in the subspace belonging to $\sum _j S_j^z=M$ 
by numerical diagonalizations 
based on the Lanczos algorithm and/or the householder algorithm. 
Here, $M$ takes an integer from zero to the saturation value 
$M_{\rm s}$ ($=S N_{\rm s}$). 
The energy is denoted by $E(N_{\rm s},M)$. 
To achieve calculations of large clusters, 
a part of Lanczos diagonalizations has been carried out 
using the MPI-parallelized code, which was originally 
developed in the study of the Haldane gaps\cite{HN_Terai}. 
The usefulness of our program was confirmed in large-scale 
parallelized calculations\cite{kgm_gap,s1tri_LRO}. 

For a finite-size system, 
the magnetization process is determined by 
the magnetization increase from $M$ to $M+1$ at the field 
\begin{equation}
h=E(N_{\rm s},M+1)-E(N_{\rm s},M),
\label{field_at_M}
\end{equation}
under the condition that the lowest-energy state 
with the magnetization $M$ and that with $M+1$ 
become the ground state in specific magnetic fields.  
When the lowest-energy state with the magnetization $M$ 
does not become the ground state in any field, 
the magnetization process around the magnetization $M$ 
is determined by the Maxwell construction. 

\section{Results and Discussion} 

\subsection{Shuriken-bonded honeycomb lattice}
First, we examine the magnetization process 
of the antiferromagnet 
on the {\it shuriken}-bonded honeycomb lattice 
shown in Fig.~\ref{fig1}(b). 
Note here that 
the treated system sizes are $N_{\rm s}=18$ and 36. 

\begin{figure}[htb]
\begin{center}
\includegraphics[width=7cm]{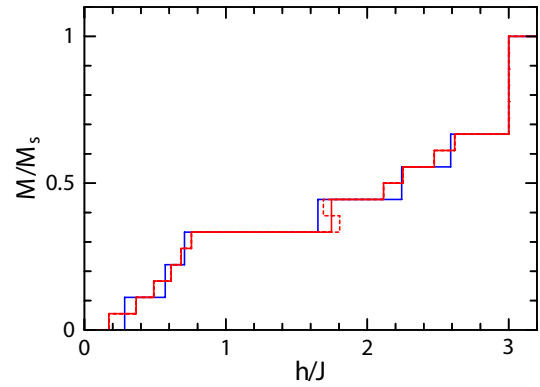}
\end{center}
\caption{(Color) Magnetization process of the antiferromagnet 
on the SBH lattice. The blue and red lines 
denote the results for $N_{\rm s}=18$ and 36, respectively. 
The solid lines represent the results after the Maxwell construction 
while the dotted line does that before the Maxwell construction. 
}
\label{fig3}
\end{figure}

Figure \ref{fig3} shows 
the results of the $T=0$ magnetization process. 
The major characteristics appear 
at the two-third and one-third heights 
of the saturation. 

At the two-third height of the saturation,  
the magnetization plateau is accompanied by 
a clear jump from this height to the saturation. 
A similar jump is known in the ordinary kagome-lattice 
and square-kagome antiferromagnets.
The mechanism of the jump 
in the kagome-lattice antiferromagnet 
is clarified by explicit eigenstates 
forming a spatially localized structure 
at hexagons of the kagome lattice discussed 
in Ref.~\ref{Zitomirsky_HTsunetsugu}. 
The same mechanism also holds in the square-kagome-lattice 
antiferromagnet; the examination was reported 
in Ref.~\ref{squagome_Richter}.  

At the one-third height of the saturation, 
a clear magnetization plateau is observed 
regardless of system size. 
An important observation is the existence 
of the magnetization jump at the higher edge 
of this height in the case of $N_{\rm s}=36$. 
The jump is not observed in the case of $N_{\rm s}=18$ 
because the resolution may be low 
owing to the smallness of the system size. 

\begin{figure}[htb]
\begin{center}
\includegraphics[width=7cm]{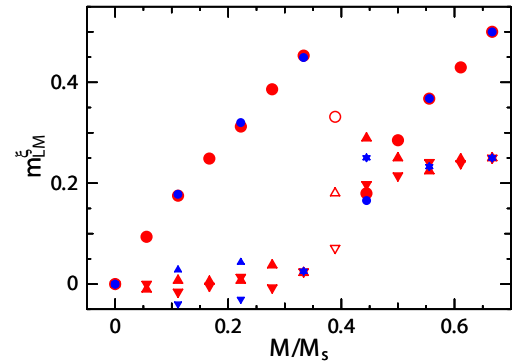}
\end{center}
\caption{(Color) Local magnetization of the state 
with respect to the global magnetization $M$ 
in the Heisenberg antiferromagnet 
on the SBH lattice.  
Blue and red symbols denote the results 
for $N_{\rm s}=18$ and 36, respectively. 
Circles, triangles, and reversed triangles 
represent the results for $\beta$, $\alpha_{12}$, 
and $\alpha_{6}$, respectively. 
Closed symbols represent data for the stably realized states, 
while open symbols denote data for the unstable states 
at the magnetization jump.  
}
\label{fig4}
\end{figure}

To clarify the relationship 
between the magnetization jump observed 
in the case of $N_{\rm s}=36$ 
and the spin-flop phenomenon, let us examine 
the local magnetization 
for both $N_{\rm s}=18$ and 36. 
The local magnetization is evaluated as 
\begin{equation}
m_{\rm LM}^{\xi} = \frac{1}{N_{\xi}} 
\sum_{j\in \xi} \langle S_j^{z} \rangle , 
\label{ave_local_mag}
\end{equation}
where $\xi$ takes $\alpha_6$, $\alpha_{12}$, and 
$\beta$. 
Here, the symbol $\langle {\cal O} \rangle$ 
denotes 
the expectation value 
of the operator ${\cal O}$ with respect 
to the lowest-energy state 
within the subspace with a fixed $M$ of interest. 
Note that the average\cite{comment_average} 
over $\xi$ is carried out 
in the case of degenerated ground states 
for some values of $M$. 
For $M$ with nondegenerated ground states, 
the results do not change regardless of 
the presence or absence of the average. 
The result is given in a plot as a function of $M$ 
in Fig.~\ref{fig4}. 
At least in the realized ground states, 
it is considered that deviations due to the 
difference in system size are small 
and that the differences between $\alpha_6$ and 
$\alpha_{12}$ are also small. 
One observes in the range 
of $0\le M \le \frac{1}{3}M_{\rm s}$
that the local magnetizations 
at $\alpha_6$ and $\alpha_{12}$ sites are maintained 
close to zero 
while the local magnetization at the $\beta$ site 
increases linearly with respect to $M$. 
At $M/M_{\rm s}=1/3$, the $\beta$ site 
reveals almost a full moment in contrast 
to the almost vanishing moments 
at the $\alpha_6$ and $\alpha_{12}$  sites.  
In the range of 
$\frac{1}{3}M_{\rm s}+2<M<\frac{2}{3}M_{\rm s}$, 
on the other hand, 
the local magnetizations 
at $\alpha_6$ and $\alpha_{12}$ sites 
are nonzero and close to the value of 
that at the $\beta$ site. 
One certainly observes that 
an abrupt change in lines along spin directions occurs 
between the regions of $M \le \frac{1}{3}M_{\rm s}$ 
and $M > \frac{1}{3}M_{\rm s}$. 
This is indeed the spin-flop phenomenon. 

Let us discuss the spin states from the above 
results of the local magnetizations. 
According to the argument of the square-kagome-lattice 
antiferromagnet\cite{shuriken_lett} 
from the viewpoint of a classical-spin system, 
we focus our attention on a local triangle, 
in which a $\beta$ site and an $\alpha_{6}$ site, and 
an $\alpha_{12}$ site are its vertices.  
It is clearly observed that, 
in the states of $M \ge \frac{1}{3}M_{\rm s} +2$, 
the local magnetizations at the $\alpha_{6}$, $\alpha_{12}$, 
and $\beta$ sites 
are nonzero but significantly smaller than the full moment. 
These results suggest that 
the spins show canting angles that are intermediate. 
Therefore, the states  $M \ge \frac{1}{3}M_{\rm s} +2$ 
suggest umbrella states. 
At $M = \frac{1}{3}M_{\rm s}$, on the other hand, 
the result that the local moment at the $\beta$ site shows 
almost a full moment indicates that the spin at the $\beta$ site 
is almost parallel to the $z$-axis. 
The vanishing moments at $\alpha_{6}$ and $\alpha_{12}$ sites 
indicate various possibilities of spin directions 
of these two sites. 
One possibility is that 
the spin at the $\alpha_{6}$ site 
and the spin at the $\alpha_{12}$ site 
are almost perpendicular 
to the $z$-axis. 
The $\alpha_{6}$ and $\alpha_{12}$ spins should be antiparallel 
to each other 
because the total magnetization may not become
polarized perpendicular to the $z$-axis 
within the consideration of a single local triangle. 
Although 
the expectation values 
$\langle S_{j}^{x} \rangle$ and 
$\langle S_{j}^{y} \rangle$ are zero 
at the $\alpha_{6}$ and $\alpha_{12}$ sites, 
the classical intuitive image 
of spin configuration at $M = \frac{1}{3}M_{\rm s}$ 
is the upside-down T structure\cite{shuriken_lett}, 
i.e., two spins at $\alpha_{6}$ and $\alpha_{12}$ sites 
are antiparallel in the $x$-$y$ plane and three 
spins at $\alpha_{6}$, $\alpha_{12}$, and $\beta$ sites 
are in the same plane. 
Another possibility is that 
one of the $\alpha_{6}$ and $\alpha_{12}$ spins 
is up 
and that 
the other spin is down 
with respect to the $z$-direction. 
In this spin configuration, the vanishing moment 
is a consequence of the average of up-spins and down-spins 
over {\it shurikens}. 
In both cases of the configurations 
in the $\alpha_{6}$ and $\alpha_{12}$ spins at $M = \frac{1}{3}M_{\rm s}$, 
the discontinuity from the states of $M \ge \frac{1}{3}M_{\rm s}$ 
can occur
because there exists a significant reduction in discontinuity 
in $m_{\rm LM}^{\beta}$ from $M = \frac{1}{3}M_{\rm s}$ 
to $M > \frac{1}{3}M_{\rm s}$.  
In the state at $M = \frac{1}{3}M_{\rm s}$, 
if one takes the quantum nature into account, 
it is considered that 
a quantum spin singlet state, 
including both the configurations 
of the $\alpha_{6}$ and $\alpha_{12}$ spins 
mentioned above, is realized 
at a local part of two $\alpha_6$-site spins 
and two $\alpha_{12}$-site spins 
forming a square inside a {\it shuriken}. 
Although there are two singlet states in 
the isolated four-spin system, 
it is an open issue as to which of the two is 
the singlet in the state at $M = \frac{1}{3}M_{\rm s}$, 
which should be tackled in future studies. 

\begin{figure}[htb]
\begin{center}
\includegraphics[width=7cm]{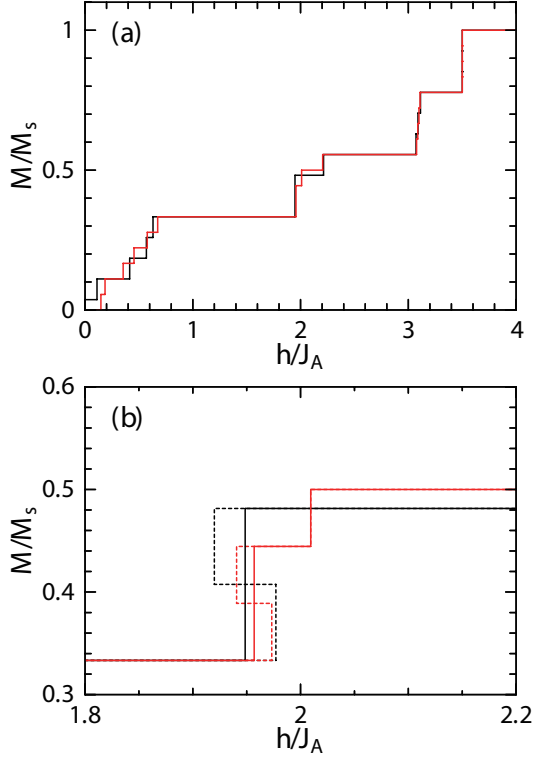}
\end{center}
\caption{(Color) Magnetization process of the antiferromagnet 
on the $\sqrt{3}\times\sqrt{3}$-distorted kagome lattice 
for $J_{\rm B}/J_{\rm A}= 1.25$: 
(a) the entire region of $0\le M \le M_{\rm s}$ and 
(b) the zoomed-in view 
at around the higher-field edge 
of the one-third height of the saturation. 
Black lines and red lines 
denote the results for $N_{\rm s}=27$ and 36, respectively. 
In (b), the solid lines represent the results obtained 
after the Maxwell construction 
while the dotted lines do those obtained 
before the Maxwell construction. 
}
\label{fig5}
\end{figure}

\subsection{Distorted kagome lattice}
Next, we examine the magnetization process 
of the antiferromagnet 
on the $\sqrt{3}\times\sqrt{3}$-distorted kagome lattice. 
We treat system sizes of $N_{\rm s}=27$ and 36, 
as shown in Figs.~\ref{fig2}(a) and \ref{fig2}(b), respectively. 
Figure \ref{fig5} shows the result 
of the $T=0$ magnetization process. 
In the present study, 
let us focus our attention 
on the behavior around the higher-field edge 
of the one-third height of the saturation
because the brief behavior in the whole range 
of the magnetic field is discussed in Ref.~\ref{Hida_kagome}. 

A magnetization jump is clearly observed 
at the one-third height 
not only for $N_{\rm s}=27$ but also for $N_{\rm s}=36$, 
although the behavior has already been detected for $N_{\rm s}=27$, 
as reported in Ref.~\ref{Hida_kagome}. 
This result regardless of the system sizes 
suggests that the jump survives in larger systems. 
It is, therefore, reasonable to consider 
that the jump is not an artifact 
owing to the finite-size effect 
but an intrinsically thermodynamic behavior. 

\begin{figure}[htb]
\begin{center}
\includegraphics[width=7cm]{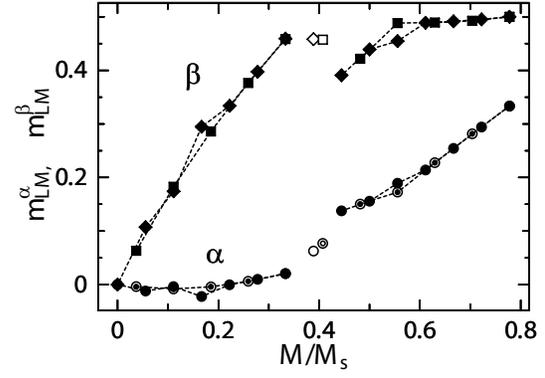}
\end{center}
\caption{
Local magnetization 
of the state with global magnetization $M$ 
in the Heisenberg antiferromagnet 
on the $\sqrt{3}\times\sqrt{3}$-distorted kagome lattice 
for $J_{\rm B}/J_{\rm A}= 1.25$. 
$m_{\rm LM}^{\alpha}$ for $N_{\rm s}=27$ and $N_{\rm s}=36$
are shown by circles and double circles, respectively.  
$m_{\rm LM}^{\beta}$ 
for $N_{\rm s}=27$ and $N_{\rm s}=36$
are shown by diamonds and squares, respectively.  
Closed symbols represent data for the stably realized states 
while open symbols denote data for the unstable states 
at the magnetization jump.  
}
\label{fig6}
\end{figure}

We examine in this subsection the behavior of $m_{\rm LM}^{\xi}$ 
given by Eq.~(\ref{ave_local_mag}), 
where $\xi$ takes $\alpha$ and $\beta$. 
The results are shown in Fig.~\ref{fig6}. 
It is a common behavior that 
a local magnetization at the $\beta$ site 
increases linearly to $M=\frac{1}{3}M_{\rm s}$ 
while that at the $\alpha$ site is maintained 
at almost zero 
in the range of $0\le M \le \frac{1}{3}M_{\rm s}$. 
In the range of  $M  \ge \frac{1}{3} M_{\rm s} +2 $, 
on the other hand, 
a local magnetization at the $\alpha$ site 
becomes nonzero, 
although the values of $m_{\rm LM}^{\xi}$ 
are different between $\alpha$ and $\beta$ sites. 
Note here that 
when data of $m_{\rm LM}^{\xi}$ 
within the range of  $M  \ge \frac{1}{3} M_{\rm s} +2 $ 
are extrapolated to $\frac{1}{3} M_{\rm s}$, 
extraplated 
results of $m_{\rm LM}^{\xi}$ 
do not seem to match the corresponding values 
of $m_{\rm LM}^{\xi}$ at $M=\frac{1}{3} M_{\rm s}$.  
This discontinuity strongly suggests 
an abrupt change in lines along spin directions 
between the regions of $M \le \frac{1}{3}M_{\rm s}$ 
and $M > \frac{1}{3}M_{\rm s}$, namely, 
the spin-flop phenomenon.  
Within the argument of classical spins 
of a local triangle, the situation is the same as 
in the case of the SBH lattice, namely,  
the spin state at $M \ge \frac{1}{3} M_{\rm s}+2$ 
can be understood on the basis of the state of the umbrella type. 
In the spin state at $M=\frac{1}{3} M_{\rm s}$, 
$\alpha$ spins form a quantum-mechanical singlet   
including two neighboring $\alpha$ spins that are 
antiparallel to each other. 
Since there is a difference in the number of $\alpha$-site spins
between the SBH lattice and 
the $\sqrt{3}\times\sqrt{3}$-distorted kagome lattice, 
the singlet is realized from six spins in the 
$\sqrt{3}\times\sqrt{3}$-distorted kagome lattice; 
the situation is more complicated. 
Although a difference exists, 
the brief behavior of spins 
in the case of the $\sqrt{3}\times\sqrt{3}$-distorted 
kagome lattice is common to the 
cases of the square-kagome lattice 
and the SBH lattice. 

\begin{table}[hbt]
\caption{
Correlation functions for $N_{\rm s}=27$ cluster 
at $J_{\rm B}/J_{\rm A}= 1.25$. 
We select a pair of sites $i$ and $j$ so that 
the distance between these sites is the largest in this cluster. }
\label{table1}
\begin{center}
\begin{tabular}{cccc}
\hline
 &     & $M=\frac{1}{3}M_{\rm s}$ & $M=\frac{1}{3}M_{\rm s}+2$ \\
\hline
$i,j\in \alpha$  
  & \multicolumn{1}{@{}c@{}}{$\begin{array}{c}
                             \langle S_{i}^{z} S_{j}^{z}\rangle  \\
                             \langle S_{i}^{x} S_{j}^{x}\rangle  \end{array}$}
  & \multicolumn{1}{@{}c@{}}{$\begin{array}{r}
                             0.00016   \\
                             0.00099   \end{array}$}
  & \multicolumn{1}{@{}c@{}}{$\begin{array}{c}
                             0.02548  \\
                             0.00642  \end{array}$}\\
\hline
$i,j\in \beta$ 
  & \multicolumn{1}{@{}c@{}}{$\begin{array}{c}
                             \langle S_{i}^{z} S_{j}^{z}\rangle  \\
                             \langle S_{i}^{x} S_{j}^{x}\rangle  \end{array}$}
  & \multicolumn{1}{@{}c@{}}{$\begin{array}{r}
                             0.21024  \\
                             -0.00157  \end{array}$}
  & \multicolumn{1}{@{}c@{}}{$\begin{array}{c}
                             0.17637  \\
                             0.01497  \end{array}$}\\
\hline
\end{tabular}
\end{center}
\end{table}

To confirm the spin direction argued above, 
let us consider the correlation functions 
$\langle S_{i}^{z} S_{j}^{z}\rangle$
and 
$\langle S_{i}^{x} S_{j}^{z}\rangle$ 
between a pair of sites $i$ and $j$.  
For such examinations based on correlation functions, 
it is desirable to observe cases when the distance 
between $i$ and $j$ at the pair is sufficiently large 
and when $i$ and $j$ are in the same group of sites, 
namely, $\alpha$ or $\beta$. 
It is difficult in such studies as the present one 
based on numerical-diagonalization calculations 
to realize a situation that these conditions are satisfied 
completely. 
Even in this situation, 
it is helpful to observe data of correlation functions 
$\langle S_{i}^{z} S_{j}^{z}\rangle$
and 
$\langle S_{i}^{x} S_{j}^{z}\rangle$ 
when $i$ and $j$ in the same group are 
the most distant in finite-size clusters. 
The $N_{\rm s}=27$ cluster is this case 
in the present study\cite{comment_N36}.  
The most distant pair in the $N_{\rm s}=27$ cluster 
is illustrated in Fig.~\ref{fig2}(a). 
Table~\ref{table1} shows results 
for the two cases of $M$ around the magnetization jump. 
At $M=\frac{1}{3}M_{\rm s}$, 
only $\langle S_{i}^{z} S_{j}^{z}\rangle$ for $i,j\in \beta$ 
is dominant, and  
the absolute values of other quantities are 
very small. 
These results agree well with 
the result of local magnetizations, 
where $m_{\rm LM}^{\beta}$ is almost saturated 
and $m_{\rm LM}^{\alpha}$ is almost vanishing at $M=\frac{1}{3}M_{\rm s}$. 
At $M=\frac{1}{3}M_{\rm s}+2$, 
on the other hand, 
$\langle S_{i}^{z} S_{j}^{z}\rangle$ for $i,j\in \beta$ 
becomes slightly smaller than that at $M=\frac{1}{3}M_{\rm s}$; 
$\langle S_{i}^{x} S_{j}^{x}\rangle$ for $i,j\in \beta$ 
increases from that at $M=\frac{1}{3}M_{\rm s}$. 
These results indicate that a $\beta$-site spin leans from the $z$-axis. 
For $i,j\in \alpha$, both 
$\langle S_{i}^{z} S_{j}^{z}\rangle$ 
and 
$\langle S_{i}^{x} S_{j}^{x}\rangle$ 
at $M=\frac{1}{3}M_{\rm s}+2$ 
become larger than those at $M=\frac{1}{3}M_{\rm s}$. 
These results suggest that 
an $\alpha$-site spin also leans from the $z$-axis. 
These leans of $\alpha$- and $\beta$-site spins 
are consistent with the state of the umbrella-type 
suggested from the discussion 
based on the local magnetizations. 
Note that these results for the single $N_{\rm s}$ 
should be reexamined in future studies 
from the viewpoint of system-size dependences. 

\begin{figure}[htb]
\begin{center}
\includegraphics[width=7cm]{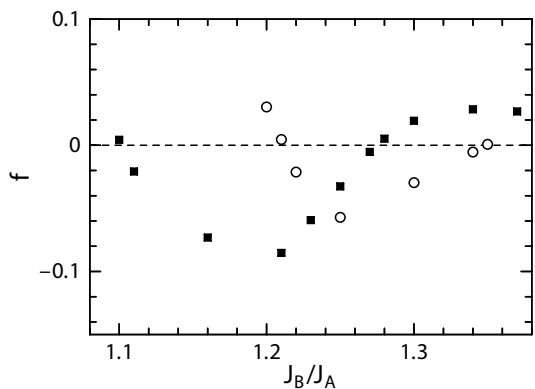}
\end{center}
\caption{
$J_{\rm B}$ dependence of $f$ at 
the one-third height of the saturation 
in the magnetization process. 
Open circles and closed squares denote 
the result of $f$ for $N_{\rm s}=27$ and $N_{\rm s}=36$, 
respectively. 
}
\label{fig7}
\end{figure}

Finally, let us investigate 
the region of an interaction parameter
where the magnetization jump appears 
in the magnetization process of the antiferromagnet 
on the $\sqrt{3}\times\sqrt{3}$-distorted kagome lattice. 
Recall here that 
it has already been pointed out that the region is narrow 
for $N_{\rm s}=27$ in Ref.~\ref{Hida_kagome}.   
To clarify the region, 
we evaluate 
\begin{equation}
f=E(N_{\rm s},M)-2E(N_{\rm s},M+1)+E(N_{\rm s},M+2), 
\end{equation}
at $M=(1/3)M_{\rm s}$. 
If the quantity $f$ is negative, 
the backbending behavior is realized 
before the Maxwell construction is carried out. 
The result is shown in Fig.~\ref{fig7}. 
The region is $1.21 \simle J_{\rm B}/J_{\rm A} \simle 1.35$ 
and $1.10 \simle J_{\rm B}/J_{\rm A} \simle 1.28$ 
for $N_{\rm s}=27$ and 36, respectively. 
The widths of the two regions are almost the same; 
it becomes larger when the system sizes are increased. 
This behavior of the width suggests that 
the region survives in the thermodynamic limit. 
Another interesting feature is that 
the region of $N_{\rm s}=36$ is closer 
to the undistorted kagome point of $J_{\rm A}=J_{\rm B}$ 
than that of  $N_{\rm s}=27$. 
At the present stage, it is difficult to 
examine what the difference in the position of the region means 
because the height of the unstable state of $M=\frac{1}{3} M_{\rm s}+1$ 
is different between $N_{\rm s}=27$ and 36.  
Recall that just at $J_{\rm A}=J_{\rm B}$, 
the behavior at the higher-field edge of the 
one-third height of the saturation in the magnetization process 
is continuous between $M=\frac{1}{3} M_{\rm s}$ and 
$M > \frac{1}{3} M_{\rm s}$ as a characteristic 
of the magnetization ramp\cite{kgm_ramp,Sakai_HN_PRBR,
HN_Sakai_PSS,HN_Sakai_SCES} 
proposed in the kagome-lattice 
antiferromagnet: the continuity is observed 
not only in the magnetization but also in its gradient. 
It is worth emphasizing that 
the region at $J_{\rm B}/J_{\rm A}\sim 1.2$ where 
the magnetization jump appears owing to the spin-flop phenomenon 
is located very close to the undistorted point $J_{\rm B}=J_{\rm A}$ 
at which the continuous behavior is realized. 
The relationship between the two cases will be clarified 
in future studies. 
Near the undistorted point $J_{\rm B}=J_{\rm A}$, 
another ferromagnetic phase of the non-Lieb-Mattis type 
is known to be 
realized\cite{collapse_ferri,Shimokawa_JPSJ} 
owing to a distortion of a different type. 
Further investigations from various viewpoints 
are required to understand much better 
the physics of the undistorted point $J_{\rm B}=J_{\rm A}$. 

\section{Conclusion and Remarks} 

We have studied the Heisenberg antiferromagnet 
on the two types of two-dimensional lattices, 
the kagome lattice with distortion 
and the {\it shuriken}-bonded honeycomb lattice, 
by the numerical diagonalization method. 
The magnetization processes of these models 
have been investigated. 
Our particular interest is the behavior 
at the one-third height of the saturation,  
where we detect the magnetization jump 
at the higher-field edge at the height. 
We successfully confirm that the jump occurs 
as a consequence of the spin-flop phenomenon 
from the analysis of local magnetizations and 
correlation functions\cite{comment_CF}.  
The present result indicates that 
such a spin-flop phenomenon is not just a rare incident 
in the square-kagome-lattice antiferromagnet 
but occurs in other various cases 
despite the fact that 
all these systems are isotropic in spin space. 
The present result also clarifies that 
not only the {\it shuriken} structure with four teeth is a trigger 
but a different {\it shuriken} structure with six teeth also 
induces the spin-flop phenomenon. 
In the magnetization process 
of the Heisenberg 
antiferromagnet on the Cairo 
pentagon lattice\cite{Ressouche_Cairo,Rousochatzakis_Cairo}, 
a similar behavior of magnetization jump appears. 
Since this lattice does not include the structure of triangles 
in it, it is an interesting example, and 
the results will be reported 
in separate papers\cite{HNakano_Cairo_lt,Isoda_Cairo_full}. 

Jumps in the magnetization process are observed in several cases 
within the isotropic quantum Heisenberg antiferromagnets. 
One is the one-dimensional case in Ref.~\ref{Honecker_Mila_Troyer}. 
The origin of the jump is also magnetic frustration; 
however, 
it is presently unclear whether a spin-flop phenomenon occurs 
because the local moment and 
other microscopic information 
about the spin state have not been investigated. 
Other cases of the magnetization jump are 
two finite-size systems of the Heisenberg clusters 
on an icosahedron\cite{Schroder_icosahedron} and 
a dodecahedron\cite{Konstantinidis_icosahedron}. 
However, one cannot increase the number of spins systematically to
that of an infinite system 
within the condition that one considers regular polyhedra. 
In this sense, it is difficult to compare the results 
of the present cases and 
the finite-size clusters on polyhedra. 
The jump of the icosahedron system appears 
only for a large spin amplitude of $S=4$, 
whereas no jump is observed in the case of $S<4$. 
Investigations of the present antiferromagnets 
of larger-$S$ spins might be helpful 
for capturing the relationship between these cases. 
It is also known that magnetization jumps appear 
below $M/M_{\rm s} \le 1/2$ 
in the magnetization process of the Shastry-Sutherland model 
in a strongly dimerized case\cite{TMomoi_KTotsuka}. 
Magnetization jumps can also appear 
when other spin-isotropic interactions are added 
to the isotropic quantum Heisenberg antiferromagnets. 
In Refs \ref{KKubo_TMomoi} and \ref{TMomoi_HSakamoto_KKen}, 
the jumps and the spin structure 
when the multiple-spin exchange interactions exist 
in addition to the Heisenberg triangular-lattice model are reported. 
Reference \ref{KPenc_NShannon_HShiba} clarified 
the appearance of the jump in 
the bilinear-biquadratic model on the pyrochlore lattice 
when the biquadratic terms are switched on. 
Thus, one finds that there are various origins 
of the magnetization jumps 
even though the system is isotropic in spin space. 

It has been clarified that 
magnetization jumps due to the spin-flop phenomenon 
occur more generally in frustrated quantum spin systems 
without anisotropy in spin space than we thought, 
even in fundamental cases composed of only Heisenberg interactions. 
The present models are typical ones among them. 
The present results strongly suggest 
a reconsideration of the mechanism 
underlying the spin-flop phenomenon. 
Further investigations 
of the spin-flop phenomenon and the magnetization jump 
would contribute much to our understanding of 
the frustration effect in magnetic materials. 

\section*{Acknowledgments}
We wish to thank 
Professors K.~Hida, M.~Isoda, Y.~Hosokoshi, T.~Ono, and 
Dr. N. Todoroki 
for fruitful discussions. 
This work was partly supported by Grants-in-Aid 
(Nos. 23340109, 23540388, and 24540348) 
from the Ministry of Education, Culture, Sports, Science 
and Technology (MEXT) of Japan. 
Nonhybrid thread-parallel calculations
in numerical diagonalizations were based on TITPACK version 2
coded by H. Nishimori. 
Some of the computations were 
performed using facilities of 
the Department of Simulation Science, 
National Institute for Fusion Science; 
Center for Computational Materials Science, 
Institute for Materials Research, Tohoku University; 
Supercomputer Center, 
Institute for Solid State Physics, The University of Tokyo;  
and Supercomputing Division, 
Information Technology Center, The University of Tokyo. 
This work was partly supported 
by the Strategic Programs for Innovative Research, 
MEXT, 
and the Computational Materials Science Initiative, Japan. 
The authors would like to express their sincere thanks 
to the staff of the Center for Computational Materials Science 
of the Institute for Materials Research, Tohoku University, 
for their continuous support 
of the SR16000 supercomputing facilities. 



\begin{thebibliography}{99}
\bibitem{comment_average}
Averaging was also carried out to obtain 
the data shown in Fig.~6 of Ref.~\ref{shuriken_lett}, 
although the authors forgot to mention it in their paper.
\bibitem{comment_N36} 
Note, on the other hand, that 
the site groups of $i$ and $j$ for 
the most distant pair in the $N_{\rm s}=36$ cluster 
are different from each other.
\bibitem{comment_CF}  
Analysis of correlation functions for the case 
of the SBH lattice has not been carried out 
in the present study because the analysis 
becomes more complicated 
owing to the fact that the SBH lattice includes 
more spin sites in its unit cell. 
The analysis will be performed in future studies.
\end{thebibliography}
\end{document}